\documentclass[prl,aps,amsmath,twocolumn,floatfix,showpacs]{revtex4}
\usepackage{epsfig}
\usepackage{bm}

\begin{document}

\title{Tomography of entangled massive particles}
\author{C.~F.~Roos, G.~P.~T.~Lancaster, M.~Riebe, H.~H{\"a}ffner,
W.~Hä{\"a}nsel, S.~Gulde, C.~Becher, J.~Eschner, F.~Schmidt-Kaler and
R.~Blatt }

\affiliation{Institut f{\"u}r Experimentalphysik, Universit{\"a}t
Innsbruck, Technikerstrasse 25, A-6020 Innsbruck, Austria}

\date{\today}

\begin{abstract}
We report on tomographic means to study the stability of a qubit
register based on a string of trapped ions.  In our experiment,
two ions are held in a linear Paul trap and are entangled
deterministically by laser pulses that couple their electronic and
motional states. We reconstruct the density matrix using single
qubit rotations and subsequent measurements with near-unity
detection efficiency. This way, we characterize the created Bell
states, the states into which they subsequently decay, and we
derive their entanglement, applying different entanglement
measures.
\end{abstract}



 \maketitle

Quantum state tomography\cite{vogel} allows the estimation of an
unknown quantum state that is available in many identical copies.
It has been experimentally demonstrated for a variety of physical
systems, among them the quantum state of a light
mode\cite{Smithey}, the vibrational state of a molecule\cite{Dunn}
and a single ion\cite{Leibfried}, and the wave packets of atoms of
an atomic beam\cite{Kurtsiefer}. Multi-particle states have been
investigated in nuclear magnetic resonance
experiments\cite{Chuang} as well as in experiments involving
entangled photon pairs\cite{White1,White2}. However, no experiment
to date has completely reconstructed the quantum state of
entangled massive particles
\cite{Hagley,Turchette,Kielpinski,FSK}. In the context of quantum
information theory\cite{Nielsen}, multi-particle entangled states
are considered to be an essential resource for processing
information encoded in quantum states (qubits). Many protocols
require the deterministic creation of entangled states and the
preservation of these states over times much longer than their
production time. For experiments aiming at entangling qubits, the
coupling of the qubits to the environment needs to be well
understood so that its effect can be minimized.

In this article, we describe the deterministic creation of all
four two-ion Bell states, i.e.
$\Psi_\pm=(|10\rangle\pm|01\rangle)\sqrt{2}$ and $\Phi_\pm =
(|11\rangle\pm|00\rangle)\sqrt{2}$, where $|x_1x_2\rangle,\;x_i
\in\{0,1\}$ represents the combined state of the two qubits. By
tomographically reconstructing the two-ion density matrix, we
fully characterize these states and determine their entanglement
by different means. We also measure the time evolution of these
states and determine the decay of their entanglement.

In our experiments, a qubit is encoded in a superposition of
internal states of a calcium ion. We use the $|S_{1/2},
m=-1/2\rangle$ ground state and the metastable $|D_{5/2},
m=-1/2\rangle$ state (lifetime $\tau \sim 1$~s) to represent the
qubit states $|1\rangle$ and $|0\rangle$, respectively. Two
$^{40}{\rm Ca}^+$ ions are loaded into a linear Paul trap having
vibrational frequencies of $(2\pi) 1.2$~MHz in the axial and
$(2\pi) 5$~MHz in the transverse directions. After 10 ms of
Doppler and sideband cooling \cite{Roos}, the ions' breathing mode
of axial vibration at $\omega_{\rm b}= (2\pi)\sqrt{3}\times
1.2$~MHz is cooled to the ground state $|0_{\rm b}\rangle$ (~99\%
occupation). Thereafter, the qubits are initialized in the
$|11\rangle$ state. For quantum state engineering, we employ a
narrowband Ti:Sapphire laser which is tightly focussed onto either
one of the two ions. By exciting the $S_{1/2}$ to $D_{5/2}$
quadrupole transition near 729 nm ($\gamma = 0.16$~Hz), we prepare
a single ion in a superposition of the $|0\rangle$ and $|1\rangle$
states. If the laser excites the transition on resonance
(''carrier transition``), the ion's vibrational state is not
affected whereas if the laser frequency is set to the transition's
upper motional sideband (''blue sideband``), the electronic states
become entangled with the motional states $|0_{\rm b}\rangle$ and
$|1_{\rm b}\rangle$ of the breathing mode. We use an
acousto-optical modulator to switch between carrier and sideband
transition frequency and to control the phase of the light field
\cite{FSK,Roos,FSK2,Haeffner}. An electro-optical beam deflector
switches the laser beam from one ion, over a distance of
5.3~$\mu$m, to the other ion within 15~$\mu$s. Directing the beam
which has a width of 2.5~$\mu$m (FWHM at the focus) onto one ion,
the intensity on the neighboring ion is suppressed by a factor of
$2.5\cdot 10^{-3}$. By a sequence of laser pulses of appropriate
length, frequency, and phase, the two ions are prepared in a Bell
state as described below. For detection of the internal quantum
states, we excite the $S_{1/2}$ to $P_{1/2}$ dipole transition
near 397~nm and monitor the fluorescence for 15~ms with an
intensified CCD camera separately for each ion. Fluorescence
indicates that the ion was in the $S_{1/2}$ state, no fluorescence
reveals that it was in the $D_{5/2}$ state. By repeating the
experimental cycle 200 times we find the average populations of
all product basis states $|00\rangle$, $|01\rangle$, $|10\rangle$
and $|11\rangle$.

We create a Bell state by applying laser pulses to ion 1 and 2 on
the blue sideband and the carrier. Using the Pauli spin matrices
$\sigma_x$,$\sigma_y$,$\sigma_z$\cite{eigenvectornote} and the
operators $b$ and $b^\dag$ that annihilate and create a phonon in
the breathing mode, we denote single qubit carrier rotations of
qubit $\alpha$ by
\begin{equation}\label{rcar}
  R_\alpha(\theta,\phi)=\exp\left[i\frac{\theta}{2}\left(
  \sigma_x^{(\alpha)}\cos\phi-\sigma_y^{(\alpha)}\sin\phi\right)\right]
\end{equation}
and rotations on the blue sideband of the vibrational breathing mode by
\begin{equation}\label{rblue}
  R_\alpha^+(\theta,\phi)=\exp\left[i\frac{\theta}{2}\left(
  \sigma_x^{(\alpha)} b^\dag\cos\phi-\sigma_y^{(\alpha)} b
  \sin\phi\right)\right].
\end{equation}
Now, we produce the Bell state $\Psi_\pm=(|10\rangle \pm
|01\rangle)/\sqrt{2}$ by the pulse sequence
$U_{\Psi_\pm}=R_2^+(\pi,\pm\pi/2)R_2(\pi,\pi/2)R_1^+(\pi/2,-\pi/2)$
applied to the $|11\rangle$ state. The pulse $R_1^+(\pi/2,-\pi/2)$
entangles the motional and the internal degrees of freedom, the
next two pulses $R_2^+(\pi, \pm\pi/2)R_2(\pi,\pi/2)$ map the
motional degree of freedom onto the internal state of ion 2.
Appending another $\pi$-pulse,
$U_{\Phi_\pm}=R_2(\pi,0)U_{\Psi\pm}$, produces the state
$\Phi_\pm$ up to a global phase. The pulse sequence takes less
than 200~$\mu$s.

To account for experimental imperfections, we describe the quantum
state by a density matrix $\rho$. For its experimental
determination we expand $\rho$ into a superposition $\rho =  \sum
\nolimits_{i} \lambda_i O_i$ of mutually orthogonal Hermitian
operators $O_i$, which form a basis and obey the equation ${\rm
tr}(O_iO_j)=4\delta_{ij}$\cite{Fano}. Then, the coefficients
$\lambda_i$ are related to the expectation values of $O_i$ by
$\lambda_i ={\rm tr}(\rho O_i)/4$. For a two-qubit system, a
convenient set of operators is given by the 16 operators
$\sigma_i^{(1)}\otimes \sigma_j^{(2)}$, $(i,j=0,1,2,3)$, where
$\sigma_i^{(\alpha)}$ runs through the set of Pauli matrices
${1,\sigma_x, \sigma_y, \sigma_z,}$ of qubit $\alpha$.

The reconstruction of the density matrix $\rho$ is now
accomplished by measuring the expectation values
$\langle\sigma_i^{(1)}\otimes \sigma_j^{(2)}\rangle_\rho$. A
fluorescence measurement projects the quantum state into one of
the states $|x_1x_2\rangle$, $x_i \in \{0,1\}$. By repeatedly
preparing and measuring the quantum state, we obtain the average
population in states $|x_1x_2\rangle$ from which we calculate the
expectation values of $\sigma_z^{(1)}$, $\sigma_z^{(2)}$ and
$\sigma_z^{(1)}\otimes\sigma_z^{(2)}$. To measure operators
involving $\sigma_y$, we apply a transformation U that maps the
eigenvectors of $\sigma_y$ onto the eigenvectors of $\sigma_z$,
i.e. $U\sigma_yU^{-1} = \sigma_z$, where $U=R(\pi/2, \pi)$.
Similarly, the operator $\sigma_x$ is transformed into $\sigma_z$
by choosing $U=R(\pi/2, 3\pi/2)$. Therefore, all expectation
values can be determined by measuring $\sigma_z^{(1)}$,
$\sigma_z^{(2)}$ or $\sigma_z^{(1)}\otimes\sigma_z^{(2)}$. To
obtain all 16 expectation values, nine different settings have to
be used (see table \ref{ReconstrPulseTable}).
\begin{table}
\caption{\label{ReconstrPulseTable} \rm{Additional pulses used for
reconstructing a two-qubit quantum state. All expectation values
needed for quantum state reconstruction can be determined by
applying either one of the nine different transformations given in
the table below and subsequently measuring the qubit state. Only
the expectation values of the operators displayed in bold face are
used to reconstruct the quantum state. }}
 \center
\begin{tabular}{|c|c|c|l|l|l|}
\hline
& \multicolumn{2}{|c|}{Transformation applied to} & \multicolumn{3}{|c|}{Measured }\\
  & ion 1 & ion 2& \multicolumn{3}{|c|}{expectation values} \\
 \hline
 1 &           - &                       - & $\bm{\sigma_z^{(1)}}$ & $\bm{\sigma_z^{(2)}}$ & $\bm{\sigma_z^{(1)} \otimes \sigma_z^{(2)}}$\\
 2 & $R(\pi/2,3\pi/2)$ &                 - & $\bm{\sigma_x^{(1)}}$ & $\sigma_z^{(2)}$      & $\bm{\sigma_x^{(1)} \otimes \sigma_z^{(2)}}$\\
 3 & $R(\pi/2, \pi)$   &                 - & $\bm{\sigma_y^{(1)}}$ & $\sigma_z^{(2)}$      & $\bm{\sigma_y^{(1)} \otimes \sigma_z^{(2)}}$\\
 4 & -                 & $R(\pi/2,3\pi/2)$ & $\sigma_z^{(1)}$      & $\bm{\sigma_x^{(2)}}$ & $\bm{\sigma_z^{(1)} \otimes \sigma_x^{(2)}}$\\
 5 &                 - & $R(\pi/2, \pi)$   & $\sigma_z^{(1)}$      & $\bm{\sigma_y^{(2)}}$ & $\bm{\sigma_z^{(1)} \otimes \sigma_y^{(2)}}$\\
 6 & $R(\pi/2,3\pi/2)$ & $R(\pi/2,3\pi/2)$ & $\sigma_x^{(1)}$      & $\sigma_x^{(2)}$      & $\bm{\sigma_x^{(1)} \otimes \sigma_x^{(2)}}$\\
 7 & $R(\pi/2,3\pi/2)$ & $R(\pi/2, \pi)$   & $\sigma_x^{(1)}$      & $\sigma_y^{(2)}$      & $\bm{\sigma_x^{(1)} \otimes \sigma_y^{(2)}}$\\
 8 & $R(\pi/2, \pi)$   & $R(\pi/2,3\pi/2)$ & $\sigma_y^{(1)}$      & $\sigma_x^{(2)}$      & $\bm{\sigma_y^{(1)} \otimes \sigma_x^{(2)}}$\\
 9 & $R(\pi/2, \pi)$   & $R(\pi/2, \pi)$   & $\sigma_y^{(1)}$      & $\sigma_y^{(2)}$      & $\bm{\sigma_y^{(1)} \otimes \sigma_y^{(2)}}$\\
\hline
\end{tabular}
\end{table}
Since a finite number of experiments allow only for an estimation
of the expectation values $\langle\sigma_i^{(1)}\otimes
\sigma_j^{(2)}\rangle_\rho$, the reconstructed matrix $\rho_{\rm
R}$ is not guaranteed to be positive semi-definite\cite{Hradil}.
Therefore, we employ a maximum likelihood estimation of the
density matrix\cite{Hradil,Banaszek}, following the procedure as
suggested and implemented in refs.~\cite{Banaszek,James}. We
parametrize the density matrix by the elements of the Cholesky
matrix T related to $\rho$ by  $\rho=T^\dagger T$. Since the
reconstructed density matrix $\rho_R$ can have negative
eigenvalues, we cannot use $\rho_R$ as a starting point for the
optimization routine. Instead, we use $\rho_P = P \rho_R P / tr(P
\rho_R P)$ where P projects onto the subspace spanned by the
eigenvectors of $\rho_R$ having non-negative eigenvalues.

\begin{figure}[t,b]
\begin{center}
\epsfig{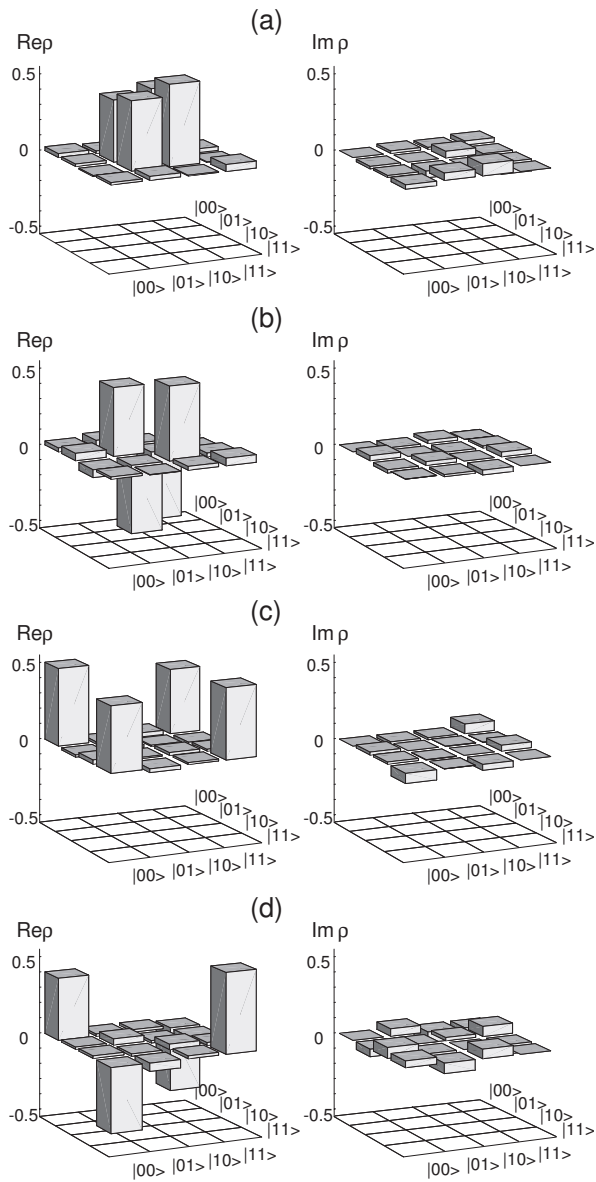}
\caption{\label{picdensmat} (a): Real and imaginary part of the
density matrix $\rho_{\Psi_+}$ that approximates
$\Psi_+=(|10\rangle+|01\rangle)/\sqrt{2}$. The measured fidelity
is $F_{\Psi_+} = \langle \Psi_+|\rho_{\Psi_+}|\Psi_+\rangle =
0.91$. (b): Real and imaginary part of the density matrix
$\rho_{\Psi_-}$ that approximates
$\Psi_-=(|10\rangle-|01\rangle)/\sqrt{2}$. The measured fidelity
is $F_{\Psi_-}= 0.90$. (c),(d): Density matrix elements of
$\rho_{\Phi_+}$ (c) and $\rho_{\Phi_-}$ (d). Here, $F_{\Phi_+} =
0.91$ and $F_{\Phi_-} = 0.88$. }
\end{center}
\end{figure}

For the pulse sequence that is designed to produce the state
$\Psi_+ = (|10\rangle+|01\rangle)/\sqrt{2}$, we obtain the density
matrix $\rho_{\Psi_+}$ shown in Fig.~\ref{picdensmat}a. The
fidelity $F$ of the reconstructed state is $F = \langle
\Psi_+|\rho_{\Psi_+}|\Psi_+\rangle = 0.91$. To produce the state
$\rho_{\Psi_-} = (|10\rangle-|01\rangle)/\sqrt{2}$, we change the
phase of the sideband $\pi$-pulse by $\pi$ and experimentally
obtain the density matrix $\rho_{\Psi_-}$ shown in
Fig.~\ref{picdensmat}b. For
$\Phi_\pm=(|11\rangle\pm|00\rangle)/\sqrt{2}$, we find the density
matrices depicted in Fig.~\ref{picdensmat}c and d.

Having reconstructed a Bell state's density matrix, we can now
check that the two qubits are indeed entangled. It has been shown
that a mixed state $\rho$ of two qubits is entangled if and only
if its partial transpose $\rho^{PT}$ has a negative
eigenvalue\cite{Peres,Horodecki}. Not surprisingly, the partial
transpose of the density matrix $\rho_{\Psi_+}$
(Fig.~\ref{picdensmat}a) has eigenvalues $\{-0.42(2), 0.40(2),
0.49(2), 0.53(3)\}$ close to the values of a maximally entangled
state $\{-0.5, 0.5, 0.5, 0.5\}$. Errors in the determination of
the density matrix elements and of quantities derived from them
occur mainly as a consequence of quantum projection noise.
Systematic effects like pulse length errors or addressing errors
(coherent excitation of an ion by stray light) play a minor role.
We estimate the magnitude of quantum projection noise by a
bootstrapping technique \cite{Efron} where the reconstructed
density matrix serves for calculating probability distributions
used in a Monte-Carlo simulation of our experiment.

Among the different measures put forward to quantify the
entanglement of mixed bipartite states, the entanglement of
formation, $E$, has the virtue of being analytically calculable
from the density matrix of a two-qubit system. For a separable
state, $E=0$, whereas $E=1$ for a maximally entangled state. Using
the formula given by Wootters\cite{Wotters}, we find
$E(\rho_{\Psi_+}) = 0.79(4)$. For the density matrices that
nominally correspond to the Bell states $\Psi_-$ and $\Phi_\pm$,
we measure $E(\rho_{\Psi_-}) =0.75(5)$, $E(\rho_{\Phi_+})
=0.76(4)$ and $E(\rho_{\Phi_-}) =0.72(5)$.

All of these Bell states violate a Clauser-Horne-Shimony-Holt
inequality. For the state $\rho_{\Psi_+}$, for example, we obtain
$|\langle A \rangle|=2.52(6) > 2$, where we have introduced the
operator $A=\sigma_x^{(1)} \otimes
\sigma_{x-z}^{(2)}+\sigma_x^{(1)}\otimes
\sigma_{x+z}^{(2)}+\sigma_z^{(1)}\otimes
\sigma_{x-z}^{(2)}-\sigma_z^{(1)}\otimes \sigma_{x+z}^{(2)}$, with
$\sigma_{x \pm z}=(\sigma_x \pm \sigma_z)/\sqrt{2}$.

\begin{figure}[t,b]
\begin{center}
\epsfig{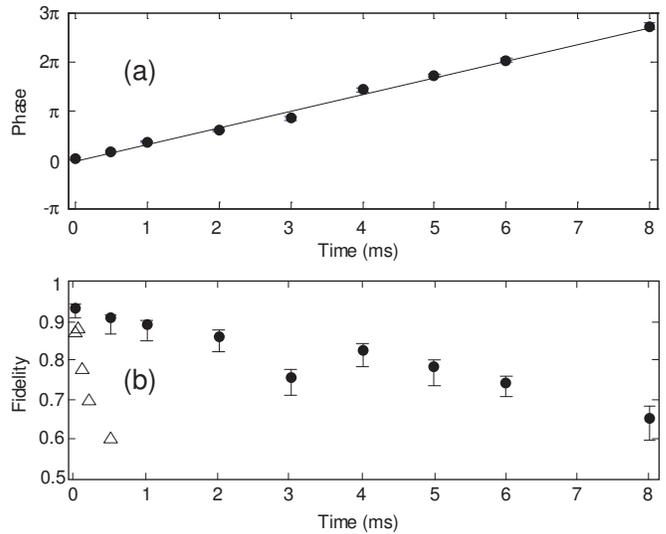}
\caption{\label{b-gradient} Time evolution of the state
$\Psi_+=(|10\rangle+|01\rangle)\sqrt{2}$. A magnetic field
gradient causes the relative phase between the $|10\rangle$ and
the $|01\rangle$ parts to evolve linearly in time. Decoherence
leads to a loss of fidelity. (a) Phase $\beta_m(t)$ for which the
overlap $F_m=
 \max_{\beta}(\langle \Psi_{\beta}|
\rho_{\Psi_+}(\tau)| \Psi_{\beta}\rangle)$ between
$\rho_{\Psi_+}(\tau)$ and states of type
$\Psi_\beta=(|10\rangle+\exp(i\beta)|01\rangle)\sqrt{2}$ is
maximized.
(b) Fidelity $F=\langle\Psi_{\beta_L}(t) |\rho(\tau)|
\Psi_{\beta_L}(\tau) \rangle)$ of the measured density matrix
$\rho(\tau)$ that nominally corresponds to the state
$\Psi_{\beta_L}(\tau)=(|10\rangle+\exp(i\beta
L)|01\rangle)\sqrt{2}$ (filled circles). The state
$\Phi_+=(|11\rangle+|00\rangle)\sqrt{2}$ (triangles) is sensitive
to fluctuations of the laser frequency and the magnetic field.
Thus, its decay occurs on a much shorter time scale. Error bars
account for quantum projection noise as well as systematic errors
in the reconstruction process. They are derived from a Monte-Carlo
simulation. }
\end{center}
\end{figure}

\begin{figure}[t,b]
\begin{center}
\epsfig{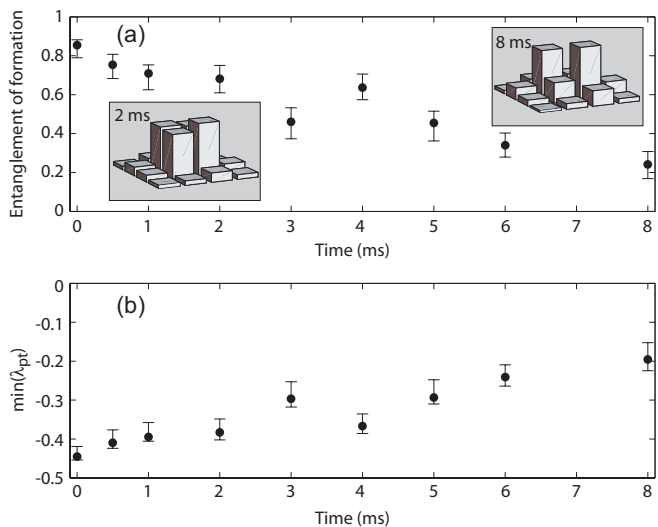} \caption{\label{decay}
Decay of the Bell state $\Psi_+=(|10\rangle+|01\rangle)\sqrt{2}$
into a statistical mixture as a function of time. (a) Entanglement
of formation and (b) smallest eigenvalue of the partial transpose
of the density matrix. The insets show the magnitude of the
density matrix elements measured after a time of 2~ms and 8~ms. }
\end{center}
\end{figure}

Once a Bell state has been produced, we monitor its evolution in
time by waiting for a time $t$ before doing state tomography.  We
expect the Bell states $\Psi_\pm$ to be immune against collective
dephasing due to fluctuations of the magnetic field or the laser
frequency\cite{Kielpinski}. However, they will only be
time-invariant if the energy separation $\hbar \omega$ between the
qubit states $|0\rangle$ and $|1\rangle$ is the same for both
qubits. A magnetic field gradient that gives rise to different
Zeeman shifts on qubits 1 and 2 leads to a linear time evolution
of the relative phase between the $|01\rangle$ and the
$|10\rangle$ component of the $\Psi_\pm$ states. This is indeed
the case in our experiments. We calculate the maximum overlap $F_m
= \max_{\beta}(\langle \Psi_{\beta}| \rho_{\Psi_+}(\tau)|
\Psi_{\beta}\rangle)$ between the density matrix
$\rho_{\Psi_+}(\tau)$ and the states
$\Psi_\beta=1/\sqrt{2}(|10\rangle+\exp(i\beta)|01\rangle)$ as a
function of time. The phase $\beta_m(\tau)$ for which the maximum
overlap $F_m$ is obtained is drawn in Fig.~\ref{b-gradient}a.
The phase changes linearly with time according to $\beta_L =
\omega_\beta \tau$ with $\omega_\beta = (2\pi) 170$~Hz, revealing
the presence of a field gradient of $dB/dz = 0.6$~G/cm in the
direction of the ion string that transforms a $\Psi_+$ state into
a $\Psi_-$ state within 3~ms and vice versa\cite{gradient}. The
decay of the Bell state into a mixed state leads to a slow decline
of the fidelity $F = \langle \Psi_{\beta_L}|  \rho_{\Psi_+}(\tau)|
\Psi_{\beta_L} \rangle$ over time which is shown in
Fig.~\ref{b-gradient}b. We find that the $\Psi_\pm$ states decay
into a statistical mixture within 5~ms whereas the $\Phi_\pm$
states already decay after about 200~$\mu$s to a fidelity of
$F=0.75$. That decay can also be seen by drawing the entanglement
of formation and the smallest eigenvalue of the partial transpose
as a function of time as shown in Fig.~\ref{decay}. In some
measurements, we found even longer lifetimes of the $\Psi_\pm$
state approaching 20~ms.

With the methods demonstrated above we have deterministically
created all four Bell states. State tomography yielded the
complete information about the two-qubit quantum states. In
future, this tomographic procedure will provide an efficient tool
for evaluating quantum gates and the effect of decoherence in
multi-qubit systems.  We gratefully acknowledge support by the
European Commission (QUEST and QGATES networks), by the Austrian
Science Fund (FWF), and by the Institut f\"ur Quanteninformation
GmbH.


\begin{thebibliography}{}
\bibitem{eigenvectornote}  Note that $|0\rangle$ and $|1\rangle$ are eigenvectors of $\sigma_z$
 corresponding to the eigenvalues $+1$ and $-1$, respectively.
\bibitem{gradient} The value of the field gradient was confirmed in an independent
measurement.
\end{thebibliography}
\end{document}